\documentstyle[prb,aps,epsf,multicol]{revtex}

\begin{document}
\draft
\preprint{{\bf ETH-TH/98-12}}

\title{Supercurrents through gated
superconductor--normal-metal--superconductor contacts:
the Josephson-transistor}

\author{Daniel\ D.\ Kuhn$^{a\,}$, Nikolai M.\ Chtchelkatchev$^{b\,}$,
  Gordey B.\ Lesovik$^{b\,}$, and Gianni Blatter$^{a\,}$}

\address{$^{a\,}$Theoretische Physik, ETH-H\"onggerberg, CH-8093
  Z\"urich, Switzerland}

\address{$^{b\,}$L. D. Landau Institute for Theoretical Physics,
  117940 Moscow, Russia}

\date{\today}
\maketitle
\begin{abstract}

  {We analyze the transport through a narrow ballistic
    super\-conductor--normal-metal--super\-conductor Josephson contact
    with non-ideal transmission at the superconductor--normal-metal
    interfaces, e.g., due to insulating layers, effective mass steps,
    or band misfits (SIN interfaces). The electronic spectrum in the
    normal wire is determined through the combination of Andreev- and
    normal reflection at the SIN interfaces. Strong normal scattering
    at the SIN interfaces introduces electron- and hole-like resonances
    in the normal region which show up in the quasi-particle spectrum.
    These resonances have strong implications for the critical
    supercurrent $I_c$ which we find to be determined by the lowest
    quasi-particle level: tuning the potential $\mu_{x0}$ to the
    points where electron- and hole-like resonances cross, we find
    sharp peaks in $I_{\rm c}$, resulting in a transitor effect. We
    compare the performance of this Resonant Josephson-Transistor
    (RJT) with that of a Superconducting Single Electron Transistor
    (SSET).}

\end{abstract}


\begin{multicols}{2}
\narrowtext


\section{Introduction}

The ability to control the supercurrent flow through narrow
super\-conductor--normal-metal--super\-conductor (SNS) contacts is not
only of scientific interest but also provides many opportunities for
applications \cite{review}. With recent progress in nanofabrication
technology it has become possible to study devices in which electrons
propagate ballistically and where the transport proceeds via few
conduction channels\cite{vanWeesWharam}. Using gated SNS
junctions\cite{Takayanagi}, the transparency of the normal region can
be manipulated: in a transparent wire, Andreev scattering at the NS
boundaries produces phase-sensitive quasi-particle levels which carry
large supercurrents; conversely, if the transmission is not ideal the
admixture of normal scattering reduces the supercurrent transport.
Accordingly, SNS junctions with a tunable transmission through the
normal part define a natural setup for a superconducting transistor
device\cite{vanHouton,Chrestin,WendinShumeiko}.

Recent interest on transport through narrow channels and quantum point
contacts concentrates on diverse phenomena such as conductance
quantization in normal constrictions
\cite{vanWeesWharam,Krans,Mueller,Scheer,Glazman}, supercurrent
quantization in superconducting SNS junctions
\cite{Takayanagi,Mueller,Beenakker,Furusaki,Chtchel}, or the
transistor effect in super-links using either gated structures
\cite{vanHouton,WendinShumeiko,Akazaki} or injection techniques
\cite{Morpurgo,Schraepers,Baselmans}. Experimentally, such junctions
are fabricated using
superconductor---semi\-conductor-hetero\-structures
\cite{vanWeesWharam,Takayanagi,Akazaki,Schraepers}, break-junctions
\cite{Krans,Mueller,Scheer}, metal-nanolithography
\cite{Morpurgo,Baselmans}, or with the use of carbon
nanotubes\cite{Kasumov}. Theoretically, transport through normal
constrictions has been studied by Glazman {\it et al.}
\cite{Glazman,GlazmanKhaetskii} within a quasi-classical description
assuming adiabatic joints between the channel and the leads. The
corresponding extension to superconducting leads by means of a
scattering matrix approach is due to Beenakker\cite{Beenakker}, see
also Ref.\ \onlinecite{Bagwell}, while Furusaki {\it et
  al.}\cite{Furusaki} proceeded with the numerical analysis of
junctions with non-adiabatic geometries and non-ideal interfaces. The
evolution of the quasi-particle spectrum and the supercurrent
quantization in a gated narrow SNS junction, as well as its
transformation into a SIS tunnel junction, has been recently described
by Chtchelkatchev {\it et al.}\cite{Chtchel} Here, we extend this
analysis to the study of SINIS junctions, where `I' stands for a
non-ideal interface between the superconducting banks S and the normal
channel N.

In an ideal SNS junction the quasi-particle spectrum is determined by
the Andreev-scattering\cite{Andreev} at the SN boundaries producing
phase-sensitive levels transporting large supercurrents\cite{Chtchel}.
The position and relative arrangement of these states strongly depends
on the chemical potential in the wire as well as on the phase
difference between the superconducting banks. The inclusion of weak
normal scattering at the SIN interface will only softly modify this
quasi-particle spectrum through the mixing of (left- and right-)
current carrying Andreev states. On the contrary, strong interface
scattering introduces electron- and hole- like resonances within the
normal region, defining a new starting point. The position of these
resonances again depends on the wire's effective chemical potential.
Tuning a pair of electron- and hole-like resonances to degeneracy
these will be mixed by the Andreev-scattering and new phase sensitive
levels are formed carrying supercurrent. This mechanism then provides
a natural setup for the implementation of a superconducting Josephson
field effect transistor\cite{vanHouton,WendinShumeiko} where the
supercurrent is switched on and off by tuning the scattering
resonances into degeneracy through the manipulation of a potential,
e.g., through external gates. Model studies of such systems for
individual resonances have been carried out recently by Wendin and
coworkers \cite{WendinShumeiko}.
\begin{figure}
  \centerline{\epsfxsize=6.7cm \epsfbox{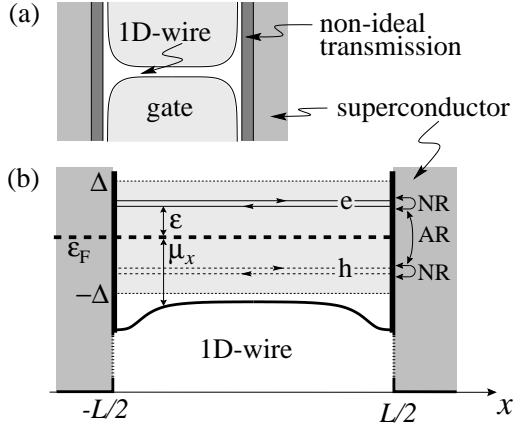}}
  \narrowtext \vspace{4mm} \caption{Narrow channel SINIS contact with
    non-ideal interfaces, e.g., insulating layers, effective mass
    steps, or band misfits. We consider adiabatic constrictions to
    avoid the mixing of transverse channels. (a) Geometrical setup
    showing the gates narrowing the wire, (b) potential landscape with
    a flat barrier bounded by $\delta$-scatterers modelling insulating
    layers.  All sources of non-ideal transmission account for {\em
      normal} reflections (NR), whereas the discontinuities of the gap
    parameter $\Delta$ are responsible for Andreev reflections (AR).}
\end{figure}
Below we proceed in two steps: After a brief definition of the problem (Sec.\
II), we first determine the scattering states of the corresponding problem where
the superconducting banks are replaced by normal-metallic leads, the NININ
junction. Second, we reinstall the superconducting banks and determine the
mixing of the (electron- and hole-like) normal scattering states through
Andreev-scattering at the normal-superconductor interface. In Sec.\ III we
analyze the quasi-particle spectrum for weak and strong normal scattering at the
NIS interfaces. Section IV is devoted to the calculation of the supercurrent; we
discuss the functionality of the Resonant Josephson Transitor (RJT) and compare
this device with the Superconducting Single Electron Transistor (SSET).

\section{Ballistic Contacts: Scattering Matrix Approach}

We consider a narrow metallic lead with few transverse channels
connecting two superconducting banks with a rectangular pair potential
$\hat\Delta(x<-L/2)=\Delta\exp(i\varphi_{\rm\scriptscriptstyle L})$,
$\hat\Delta(|x|<L/2)=0$ and $\hat\Delta(x>L/2)=\Delta
\exp(i\varphi_{\rm \scriptscriptstyle R})$, see Fig.~1(a). Joining the
channel adiabatically to the superconducting banks, the transverse
channels in the wire are separable\cite{GlazmanKhaetskii} and for each
of them the quasi-particle spectrum $\varepsilon_\nu$ is determined
through the 1D Bogoliubov-de Gennes equation (we choose states with
$\varepsilon_\nu\geq 0$)
\begin{eqnarray}
\left[ \begin{array}{cc}
  \!{\cal H}_0 & \!\! \hat\Delta(x) \! \\
  \!\! \hat\Delta^*(x) & \!\!-{\cal H}_0 \!
  \end{array}\right]
  \! \left[\begin{array}{c}\! u_\nu(x)\! \\
  \! v_\nu(x) \! \end{array} \right] = \varepsilon_\nu
  \! \left[\begin{array}{c}\! u_\nu(x)\! \\
  \! v_\nu(x) \! \end{array} \right] ,
\end{eqnarray}
with ${\cal H}_0 = -\hbar^2 \partial_x [1/2m(x)] \partial_x + U(x) - \mu_x(x)$
and where $u_\nu$ and $v_\nu$ denote the electron- and hole-like components of
the wave function $\Psi_\nu$. The potential $U(x)$ induces normal scattering at
the NS interface and is due to an insulating layer or a band offset, for
example.  Similarly, the mass $m(x)$ may change at the NS interface, again
generating normal scattering. The transverse energy $\varepsilon_\perp(x)$ of
the channel is enclosed in the effective chemical
potential\cite{GlazmanKhaetskii} $\mu_x(x)=\varepsilon_{\rm\scriptscriptstyle F}
-\varepsilon_\perp(x)$. We will also make use of the kinetic energies
$E=\varepsilon_{\rm\scriptscriptstyle F}\pm\varepsilon$ of electron ($+$) and
hole ($-$) states as measured with respect to the band bottom in the
superconductors.

The spectrum splits into continuous and discrete contributions and we
will concentrate on the latter part with $\varepsilon_\nu<\Delta$ in
the following, as it provides the main contribution to the critical
supercurrent in the most interesting transport regimes, see below.  We
solve the Bogoliubov-de Gennes equation for the SINIS junction with
the help of the usual transfer-matrix technique\cite{RiccoAzbel}. In
order to do so we first have to determine the resonance structure of
the related NININ problem that arises if we replace the
superconductors by normal metallic leads.

\subsection{NININ-junctions}

We expand the scattering states in a basis of in- and outgoing states
$\Psi_{\rm\scriptscriptstyle R,L}^{\rm\scriptscriptstyle in,out}$ on
both sides of the wire, with phases that vanish at the
interfaces\cite{conv} $\pm L/2$,
\begin{eqnarray}
\Psi_{\rm\scriptscriptstyle L}(x) &=& a_{\rm\scriptscriptstyle
L}^{\rm\scriptscriptstyle in} e^{ik(x+L/2)} +a_{\rm\scriptscriptstyle
L}^{\rm\scriptscriptstyle out} e^{-ik(x+L/2)},
\\
\Psi_{\rm\scriptscriptstyle R}(x) &=& a_{\rm\scriptscriptstyle
R}^{\rm\scriptscriptstyle in} e^{-ik(x-L/2)} +a_{\rm\scriptscriptstyle
R}^{\rm\scriptscriptstyle out} e^{ik(x-L/2)},
\end{eqnarray}
with $k(E)=\sqrt{2mE}/\hbar$ the wave vector of an incident particle.
The energy dependent scattering matrix ${\cal S}$ connects the
expansion coefficients $a_{\rm\scriptscriptstyle
  R,L}^{\rm\scriptscriptstyle in,out}$ (here, $t$ and $r$ denote the
moduli of the matrix elements),
\begin{eqnarray}
\left[\begin{array}{c}\! a_{\rm\scriptscriptstyle R}^{\rm\scriptscriptstyle
        out}\!  \\\!  a_{\rm\scriptscriptstyle L}^{\rm\scriptscriptstyle
        out}\!  \end{array} \right] = \left[ \begin{array}{cc}\!
  t\exp(i\chi^t)\!  &\!  -r\exp(i[2\chi^t-\chi^r])\!  \\\! r\exp(i\chi^r)\!
  &\!  t\exp(i\chi^t)\! \end{array}\right] \left[\begin{array}{c} \!
  a_{\rm\scriptscriptstyle L}^{\rm\scriptscriptstyle in}\! \\\!
  a_{\rm\scriptscriptstyle R}^{\rm\scriptscriptstyle in}\! \end{array}\right];
\label{scatamp}
\end{eqnarray}
with this definition of the ${\cal S}$-matrix and the basis states,
the scattering phases account for the propagation through the wire.
E.g., for a vanishing transverse energy in the normal region and in
the absence of any interface barrier a finite phase $\chi^t=kL$ is
picked up.

We describe the effective chemical potential in the normal region
through a smooth function characterized by its minimum
$\mu_x(0)=\mu_{x0}$ and a positive curvature
$m\Omega^2=\partial_x^2\mu_x$ (see Fig.~1(b)). The parameter
$\mu_{x0}$ (i.e., the diameter of the metallic wire) is assumed to be
tunable by means of external electrostatic gates. In a long wire the
potential is flat, $\hbar\Omega\ll\Delta$, and produces a sharp
switching between transmission and reflection within the energy
interval $\hbar\Omega$; the corresponding transverse energy
$\varepsilon_\perp(x)$ defines a smooth and flat potential barrier in
the interval $(-L/2,L/2)$. In the following we refer to such barriers
as {\em smooth} and $t_i\exp(i\chi^{t_i})$ and $r_i\exp(i\chi^{r_i})$
denote the associated `inner' scattering amplitudes describing the
motion of the quasi-particles between $-L/2+0$ and $L/2-0$ (we use an
analoguous definition of the `inner' scattering amplitudes as in
(\ref{scatamp})). The particular smooth geometry of such inner
barriers justifies the application of the Kemble formula\cite{Glazman}
for the transmission probability, $t_i^2=1/\{1+\exp[-2\pi(\mu_x(0)\pm
\varepsilon)/\hbar\Omega]\}$, whereas the quasi-classic method can be
used to determine the scattering phases $\chi^{t_i}$ and $\chi^{r_i}$.
\begin{figure}
  \centerline{\epsfxsize=6.5cm \epsfbox{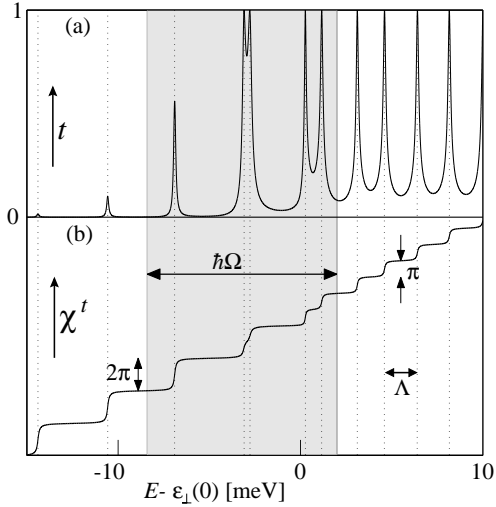}}
  \narrowtext\vspace{4mm} \caption{Modulus and phase of the
    transmission amplitude through a parabolic barrier with
    $\delta$-scatterers at the edges (parameters: $Z=2$,
    $\varepsilon_\perp(0)=5$~eV, $L=2500$~nm). Note that the energy
    $E$ is swept and $\varepsilon_\perp(0)$ is kept fixed. A similar
    resonance structure is found for other sources of non-ideal
    transmission at $\pm L/2$.}
\end{figure}
In order to obtain the global scattering amplitudes $t\exp(i\chi^{t})$
and $r\exp(i\chi^{r})$ of the junction we have to include effects of
normal scattering at the boundaries. As a typical example we consider
symmetric smooth barriers of height
$\varepsilon_{\rm\scriptscriptstyle F}-\mu_{x0}$ bounded by Dirac
scatterers of strength $V_0$ at the interfaces (such a setup describes
an NININ junction with insulating layers at the interfaces; see the
Appendix for a brief discussion of other typical cases).

First, we concentrate on the global {\it transmission} amplitude
$t\exp(i\chi^{t})$ which is easily expressed in terms of the
scattering amplitudes of the inner barrier and the Dirac
$\delta$-scatterers, see the Appendix for details of the calculation.
The main effect of the scattering at the channel boundaries is to
introduce resonances at energies $E_{\rm res}$ which we call {\em
  perfect} ({\em imperfect}) if $t(E_{\rm res})=1$ ($<1$). For
energies $E>\varepsilon_\perp(0) +\hbar\Omega/2$ the electrons easily
propagate through the channel, while the reflection at the boundaries
produces almost equidistant resonances with a spacing $\Lambda \approx
\pi/\partial_E \chi^{t_i}\sim 2(\mu_{x0}\varepsilon_{\rm
  \scriptscriptstyle L})^{1/2}$, with $\varepsilon_{\rm
  \scriptscriptstyle L} = \hbar^2\pi^2/2mL^2$ (see Fig.~2; the last
relation describes the case of a flat inner barrier). On the contrary,
for small energies below $\varepsilon_\perp(0)-\hbar\Omega/2$,
tunneling suppresses the propagation through the channel and only
imperfect resonances separated by $\sim 2\Lambda$ survive. In the
intermediate region, pairs of perfect resonances collapse and become
imperfect. Moreover, the scattering phase $\chi^t$ grows with the
energy and picks up a phase of $\pi$ and $2\pi$ at perfect and
imperfect resonances, respectively. We refer to resonances at energies
$E_{\rm res}= \varepsilon_{\rm\scriptscriptstyle
  F}+\hat\varepsilon_{\rm res}$ above
$\varepsilon_{\rm\scriptscriptstyle F}$ as {\em electronic} and to
those below, at $E_{\rm res}=\varepsilon_{\rm\scriptscriptstyle
  F}-\check \varepsilon_{\rm res}$, as {\em hole-like} resonances.
Below, we will be mostly concentrating on the propagating regime with
$E>\varepsilon_\perp(0) +\hbar\Omega/2$. In this regime the global
transmission amplitude takes the form ($t_{i\pm}\approx 1$)
\begin{eqnarray}
\label{t_delta}
t\exp(i\chi^t)=\frac{e^{i\chi^{t_i}}}{1-Z^2+i2Z+Z^2e^{i2\chi^{t_i}}},
\end{eqnarray}
with $Z=mV_0/\hbar^2k_{\rm\scriptscriptstyle F}$ the dimensionless
parameter giving the strength of the scattering potential. Resonances
of width
\begin{equation}
\Gamma = \frac{2}{\partial_E\chi^t} = \frac{2\Lambda}{\pi}\frac{T}{2-T} \qquad
{\rm with} \qquad \Lambda=\frac{\pi}{\partial_E\chi^{t_i}}
\end{equation}
and a transmission probability $T=1/(1+Z^2)$ appear as the denominator in
(\ref{t_delta}) touches the complex unit circle (note that $\Gamma
\longrightarrow 2\hbar v/L$, with $v$ the particle velocity, in the absence of
any scattering potential).

For symmetric barriers the {\it reflection} amplitude
$r\exp(i\chi^{r})$ is determined (up to $\pm1$) by the unitarity of
the scattering matrix, i.e., $r=(1-t^2)^{1/2}$ and
$\chi^r=\chi^t+\pi(n+1/2)$. The integer $n$ jumps by unity at perfect
resonances.

\subsection{SINIS-junctions}

After evaluating the normal scattering amplitudes we can reinstall the
superconductors and match the scattering states in the normal region with the
evanescent modes in the superconducting banks. We make use of the Andreev
approximation\cite{Andreev} and obtain the quantization
condition\cite{Beenakker,Chtchel,GogadzeKosevich}
\begin{eqnarray}
\label{qc} \cos(\chi_+^t-\chi_-^t-\alpha)=r_+r_-\cos\beta+t_+t_-\cos\varphi,
\end{eqnarray}
where the $+(-)$ signs refer to the kinetic energies
$\varepsilon_{\rm\scriptscriptstyle F}\pm\varepsilon$ of the
electron(hole)-like states\cite{sepNS}. The Andreev scattering at the
NS boundaries introduces the phase $\alpha=2\arccos(
\varepsilon/\Delta)$ decreasing from $\pi$ at $\varepsilon=0$ to $0$
at the gap $\varepsilon=\Delta$, as well as the phase difference
$\varphi=\varphi_{\rm\scriptscriptstyle
  L}-\varphi_{\rm\scriptscriptstyle R}$ between the two
superconducting banks. For symmetric barriers the phase
$\beta=(\chi^t_+-\chi^r_+)-(\chi^t_--\chi^r_-)$ is a multiple of $\pi$
and produces a smooth function $r_+r_-\cos\beta$ changing sign at
perfect resonances [see Fig.\ 5(b)].

The case of an ideal SNS junction where the effective chemical
potential joins smoothly to the band bottom in the superconductors has
been analyzed by Chtchelkatchev {\it et al.}\cite{Chtchel} In this
special situation the global and inner scattering amplitudes coincide
and resonances are absent. Sweeping $\mu_{x0}$, they find a one
parametric ($\mu_{x0}$) family of discrete spectra describing the
transition of an insulating SIS tunnel junction (at
$\mu_{x0}\ll-\Delta$) into a ballistic SNS structure (at
$\mu_{x0}\gg\Delta$), see Fig.~3(a). More precisely, electronic levels
with an exponentially small dependence on the phase $\varphi$ are
converted into phase sensitive Andreev-levels as $\mu_{x0}$ is
increased, i.e., as the band bottom in the wire is lowered. It turns
out that the critical supercurrent is carried by the lowest state,
$I_c=\max_\varphi[(2e/\hbar)\partial_\varphi\varepsilon_0]$, and is
realized for a phase $\varphi = \pi-0$.  Increasing the channel width,
the critical supercurrent increases in steps of
$e/(\tau_0+\hbar/\Delta)$ as new transverse channels open. The travel
time $\tau_0$ of the quasi-particles is easily calculated within the
quasi-classical scheme and approaches the asymptotic value $\tau_0
\sim L/v_{{\rm\scriptscriptstyle F},x}$ in the open channel. This
analysis explains the dependence of the non-universal critical
supercurrent steps on the junction parameters.

In the following we go beyond the analysis of Chtchelkatchev {\it et
  al.} and determine the spectrum and the supercurrent transport in a
narrow ballistic SINIS Josephson junction including resonant barriers
in the normal region, see Fig.~1(b).

\section{Quasi-particle Spectrum}

Below we shall see that the bound state spectrum of junctions with smooth
barriers is only slightly modified by switching on weak resonances, see
Fig.~3(b), with the degeneracy of the Andreev levels at $\varphi=0$ and $\pi$
lifted. For a fixed pair of levels $\varepsilon^{(\pm)}$, this splitting
$\delta\varepsilon=\varepsilon^{(+)}-\varepsilon^{(-)}$ is roughly a periodic
function of $\mu_{x0}$ with a period $\Lambda$ reflecting the resonance spacing.
As the strength of the interface scattering increases the discrete spectrum is
gradually distorted (see Fig.~3(c)); the phase sensitivity at large chemical
potential $\mu_{x0}$ (region I) is reduced and the bound state energies align
with the resonance energies of the normal NININ setup. The effect of interface
scattering on the continuum part of the spectrum has been analyzed in Ref.\
\onlinecite{GogadzeKosevich}.

\subsection{Weak Resonances}

We first concentrate on weak resonances characterized by a small
scattering parameter $Z\ll 1$. In particular, we aim at estimating the
level splitting in the Andreev spectrum shown in Fig.~3(b). Using
Eq.~(\ref{t_delta}) we evaluate the terms on the right side of the
quantization condition (\ref{qc}) to lowest order in $Z$ and find
$t_+t_-=1-2Z^2[\cos^2\chi_+^{t_i}+\cos^2\chi_-^{t_i}]$ and
$r_+r_-=4Z^2|\cos\chi_+^{t_i}||\cos\chi_-^{t_i}|$.  For symmetric
barriers and as a consequence of the unitarity of the scattering
matrix, the rescaled phase $\beta(\varepsilon)/\pi$ is an integer
valued function starting from $0$ at $\varepsilon=0$ and jumping by
unity whenever $\varepsilon$ lines up with a (perfect) resonance at
$\hat\varepsilon_{\rm res}$ or $\check\varepsilon_{\rm res}$ of the
junction.  We then may write $\cos\beta=(-1)^n$, where
$n(\varepsilon)$ denotes the number of perfect resonances within the
interval $(0,\varepsilon)$. Hence, $\cos\beta$ generates a smooth
contribution $r_+r_-\cos\beta=4Z^2\cos\chi_+^{t_i}\cos\chi_-^{t_i}$,
which is small, of order $Z^2$. Finally, the scattering phase
$\delta\chi^t=\chi_+^t-\chi_-^t$ in (\ref{qc}) may be identified with
$\delta\chi^{t_i}=\chi_+^{t_i}- \chi_-^{t_i}$; the deviation of
$\delta\chi^t$ from $\delta\chi^{t_i}$ provides only small corrections
of order $Z^2$ to the splitting in the Andreev spectrum and thus can
be disregarded.
\begin{figure}
  \centerline{\epsfxsize=6cm \epsfbox{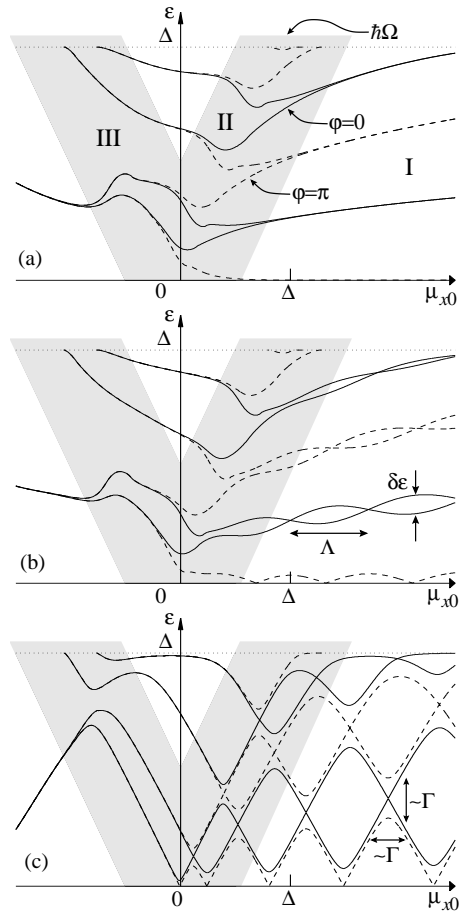}}
  \narrowtext\vspace{4mm} \caption{Discrete energy spectrum for a
    smooth parabolic potential barrier bounded by Dirac scatterers of
    strength $V_0$ ($Z = mV_0/\hbar^2k_{\rm F}$; parameters are chosen
    to emphasize the overall structure of the spectrum). In regions~I
    and II the quasi-particle energies depend on the phase $\varphi$
    ($\varphi=0$: solid lines, $\varphi=\pi$: dashed lines).  Note
    that the degeneracies in region~I are lifted for $\varphi\neq0$,
    $\pm\pi$. (a) The unperturbed spectrum for $Z=0$ as discussed by
    Chthelkatchev {\it et al.}\cite{Chtchel} (b) At $Z=0.1$ the
    junction exhibits weak resonances which lead to a splitting in the
    Andreev spectrum; the period $\Lambda$ reflects the resonance
    spacing. (c) For $Z=1$ the junction develops distinct resonances
    to which the quasi-particle levels are pinned. The remaining
    degeneracies are a consequence of the almost identical resonance
    shapes.}
\end{figure}
At $\varphi=0$ the right hand side of (\ref{qc}) can be expressed as
$R=1-2 Z^2(1-\cos\delta\chi^{t_i})(1-\cos\Sigma\chi^{t_i})$ while the
left hand side takes the form $L=\cos(\delta\chi^{t_i}-\alpha)$. The
sum $\Sigma\chi^{t_i}= \chi^{t_i}_++\chi^{t_i}_-$ of the scattering
phases depends only weakly on $\varepsilon$ but increases linearly in
$\mu_{x0}$ with a slope $2\pi/\Lambda$, as follows from linearizing
$\chi^{t_i}_{\pm}(\varepsilon,\mu_{x0})$. On the other hand, the phase
difference $\delta\chi^{t_i}$ increases with energy $\varepsilon$ but
is roughly independent of $\mu_{x0}$. Both terms $L$ and $R$ then show
an oscillating dependence on $\delta\chi^{t_i}$ and on the energy
$\varepsilon$ (see Fig.~4), while their relative phase
$\alpha(\varepsilon)$ slowly decreases from $\pi$ at $\varepsilon=0$
to $0$ at $\varepsilon=\Delta$. Thus, for small energies
($\varepsilon\ll\Delta$) the maxima in $L$ coincide with the minima in
$R$ and we obtain a large splitting,
\begin{eqnarray}
\label{gap} \delta\varepsilon\approx \frac{4}{\pi}Z\Lambda|\sin\chi^{t_i}_+|,
\end{eqnarray}
to lowest order in $Z$ (with $\chi_+^{t_i}$ evaluated at
$\varepsilon=0$).  The corresponding splitting for $\varphi=\pm \pi$
is proportional to $|\cos\chi^{t_i}_+|$ with the same prefactor. With
increasing energy the phases of $L$ and $R$ match up and the splitting
becomes small, see Fig.~4. The monotonous decrease of the splitting is
most prominent in long junctions with many trapped levels.

As manifested in (\ref{gap}) the splitting vanishes at specific
degeneracy points: we define the $(l,m)$-degeneracy via the condition
$\varepsilon_{\rm\scriptscriptstyle F}=(E^l_{\rm res}+E^m_{\rm
  res})/2$, where $l$ and $m$ count the perfect resonances. This
implies $|E^l_{\rm res}-E^m_{\rm res}|/2=\hat \varepsilon_{\rm
  res}=\check \varepsilon_{\rm res}=\varepsilon^{\rm d}_{\rm res}$.
The ideal degeneracies in (\ref{gap}) originate from the assumption
that all perfect resonances have the same symmetric shape, $t(E^l_{\rm
  res}+\delta E)=t(E^m_{\rm res}\pm\delta E)$.  For $\varphi=0$
($\pm\pi$) and for even (odd) values of $n(\varepsilon^{\rm d}_{\rm
  res})$ the right hand side of the quantization condition (multiplied
by $-1$) equals unity, $r_+r_-+t_+t_-=r_+^2+t_+^2=1$, and thus the
trapped levels appear in degenerate pairs. The degeneracy is lifted,
if we take into account that the perfect resonances are not identical.
\begin{figure}
  \centerline{\epsfxsize=7cm \epsfbox{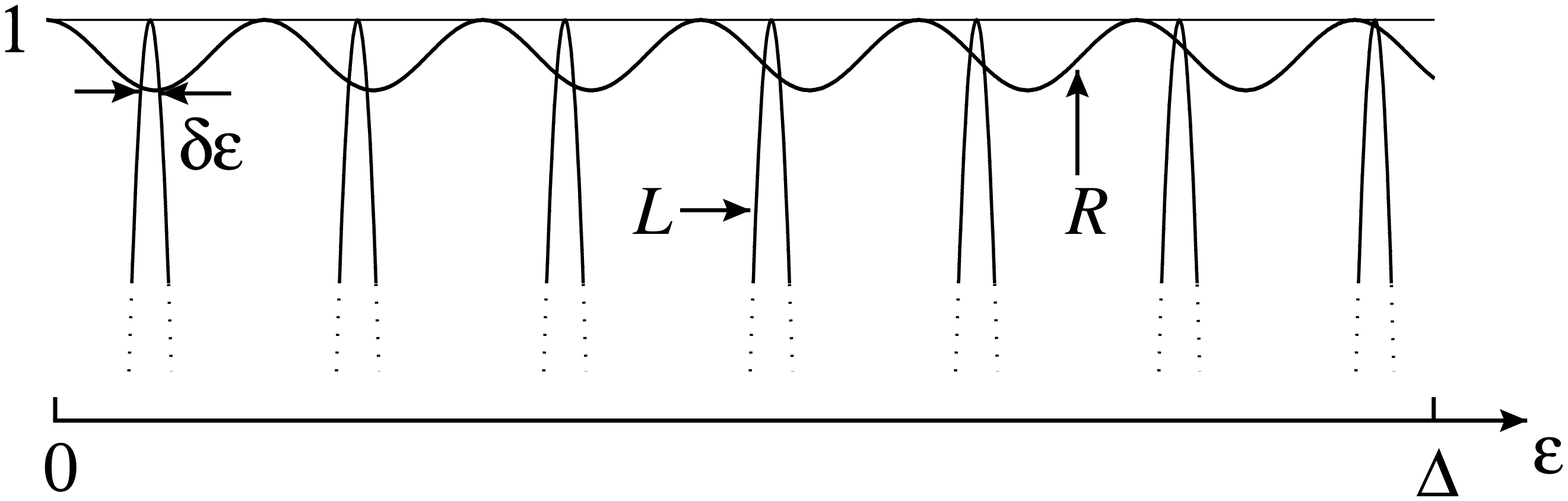}}
  \narrowtext\vspace{4mm} \caption{The quasi-particle energies are
    determined by the intersections of the right and the left hand
    side ($R$ and $L$) of the quantization condition (\ref{qc}). Every
    maximum in $L$ contributes a pair of trapped levels split by
    $\delta\varepsilon$.}
\end{figure}

\subsection{Strong Resonances}

Let us next study the effects of {\em strong} resonances. Without loss of
generality we again discuss the symmetric barrier bounded by $\delta$-scatterers
(see also Ref.\ \onlinecite{GogadzeKosevich}) . We concentrate on the open
channels in region I (see Fig.\ 3(a)); the other regimes in the
$(\varepsilon,\mu_{x0})$-plane are less interesting and we will briefly discuss
them at the end. For sharp resonances with $\Gamma \ll \Delta$ the
transmittivity close to a resonance $\varepsilon_{\rm res}$ assumes a Lorentz
profile\cite{gamma}
$t_\pm(\varepsilon)=(\Gamma/2)/\sqrt{(\varepsilon-\varepsilon_{\rm
    res})^2+(\Gamma/2)^2}$ and the scattering phase takes the usual
form $\chi^t_\pm(\varepsilon)=\chi^t(E_{\rm
  res})\pm\arctan[2(\varepsilon- \varepsilon_{\rm res})/\Gamma]$.

Fig.~5 shows the terms entering the quantization condition (\ref{qc}).
Particles incident from the left on the normal region are reflected
back ($r_+\approx 1$) unless their energy coincides with an electronic
resonance energy ($r_+(\hat\varepsilon_{\rm res})=0$ for perfect
resonances). Similarly, holes are only transmitted if their energy
corresponds to a hole-like resonance energy
($r_-(\check\varepsilon_{\rm res})=0$). Consequently, the product
$r_+r_-$ remains close to unity but sharply drops to zero at perfect
resonance energies (see Fig.~5(a)). Similar to the case of weak
resonances the product $r_+r_-$ becomes a smooth function of
$\varepsilon$ when combined with $\cos\beta=(-1)^n$, see Fig.~5(b).
Furthermore, the phase sensitive term $t_+t_-\cos\varphi$ vanishes
practically over the entire interval $[0,\Delta]$ (see Fig.~5(c)) with
the exception of specific points where particle- and hole- resonances
become {\em degenerate}, $\hat\varepsilon_{\rm res} =
\check\varepsilon_{\rm res}=\varepsilon^{\rm d}_{\rm res}$.
\begin{figure}
  \centerline{\epsfxsize=8.5cm \epsfbox{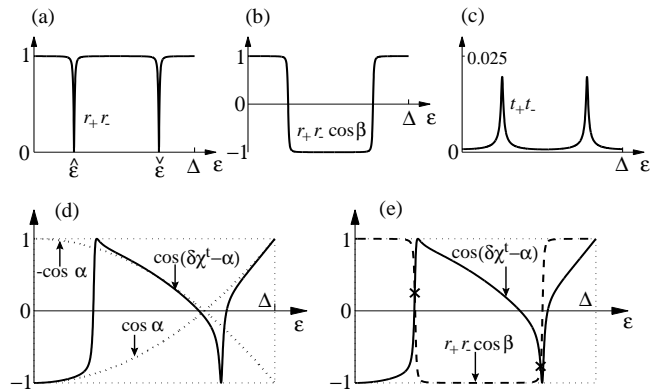}}
  \narrowtext\vspace{4mm} \caption{The various terms entering the
    quantization condition (\ref{qc}) in the limit $\Gamma\ll\Delta$:
    We show the case of perfect {\em isolated} resonances
    ($|\hat\varepsilon-\check\varepsilon|\gg\Gamma$), where the
    product $t_+t_-$ is small at all energies (here, $t_+t_-<0.025$).
    The bound state energies (marked with crosses in (e)) are
    determined by the intersections of $\cos(\delta\chi^t-\alpha)$ and
    $r_+r_-\cos\beta$.}
\end{figure}
The left hand side of (\ref{qc}) is mainly determined by the phase
difference $\delta\chi^t$ exhibiting sharp steps by $\pi$ at the
resonance energies $\check\varepsilon_{\rm res}$ and
$\hat\varepsilon_{\rm res}$ while remaining constant in between (in
our discussion of the situation away from region I below, we will have
to distinguish perfect from imperfect resonances, as the latter
involve phase jumps by $2\pi$). Hence, in the limit $\Gamma
\rightarrow 0$ we obtain $\cos(\delta\chi^t-\alpha)= (-1)^n\cos\alpha$
while a small finite value of $\Gamma$ will smooth the discontinuities
(see Fig.~5(d)). All the terms entering the quantization condition
(\ref{qc}) then are pronounced functions of $\varepsilon$ near the
resonance energies while staying roughly constant everywhere else. As
a consequence, the intersections of the left and the right hand side
of (\ref{qc}) come to lie close to the resonance energies of the
NININ-junction as shown in Fig.~5(e) and we conclude that the normal
state resonances attract the quasi-particle bound states,
$\varepsilon\approx\varepsilon_{\rm res}$ to zeroth order in $\Gamma$.
The predicate {\em electronic-} and {\em hole-like} states then can be
naturally assigned to the trapped levels in the SINIS junction, too.

In region~I (see Fig.\ 3(a)), the bound state energy close to an {\it
  isolated resonance} is determined by the implicit equation
\begin{eqnarray}
\label{level1} \varepsilon=\varepsilon_{\rm res}-
\frac{\Gamma}{2}\cot[\alpha(\varepsilon)/2].
\end{eqnarray}
This equation follows directly from the quantization condition
(\ref{qc}) with $t_+\exp(i\chi^t_+)$ described by an ideal Lorentzian
resonance while $t_-=0$, $r_-=1$, and
$\chi_-^t=\chi_+^t(\varepsilon_{\rm res}) +n(\varepsilon_{\rm
  res})\pi$ is constant (or vice versa $+\leftrightarrow -$).  For
$\varepsilon\ll\Delta$ we may approximate
$\alpha(\varepsilon)\approx\pi$, while at $\varepsilon\lesssim\Delta$
(\ref{level1}) is easily analyzed graphically (note, however, that the
above assumptions for $t_\pm$ and $\chi_\pm^t$ are only valid if
$|\varepsilon-\varepsilon_{\rm res}|\ll\Delta$). The phase
independence of the quasi-particle states in the presence of strong
normal scattering is made explicit in the result (\ref{level1}).

Close to the {\it particle-hole degenerate resonances} the phase
sensitive term $t_+t_-=(\Gamma/2)^2/ [(\varepsilon-\varepsilon^{\rm
  d}_{\rm res})^2+ (\Gamma/2)^2]$ is of order unity and we obtain
pairs of trapped levels
\begin{eqnarray}
\label{level2}
\varepsilon^{(\pm)}\! = \!\!\left\{ \begin{array}{cc}\!\!
   \varepsilon_{\rm res}^{\rm d}
    -\frac{\Gamma}{2} \! \left(\cot(\frac{\alpha}{2}) \pm\frac{|\sin(\varphi/2)|}
      {\sin(\alpha/2)} \right)\!, &\!\! n(\varepsilon_{\rm res}^{\rm d})\
    \mbox{even,}\\ \!\!
   \varepsilon_{\rm res}^{\rm d}
    -\frac{\Gamma}{2} \! \left(\cot(\frac{\alpha}{2}) \pm \frac{|\cos(\varphi/2)|}
      {\sin(\alpha/2)} \right)\!, &\!\! n(\varepsilon_{\rm res}^{\rm d})\
    \mbox{odd}\end{array} \right.
\end{eqnarray}
(we assume both scattering amplitudes $t_\pm\exp(i\chi^t_\pm)$ to be
described by the same Lorentzian centered around $\varepsilon_{\rm
  res}^{\rm d}$; Eq.\ \ref{level2} agrees with the result of Ref.\
\onlinecite{WendinShumeiko} obtained for a short junction\cite{gamma}, see also
Ref.\ \onlinecite{GogadzeKosevich}). The levels (\ref{level2}) are manifestly
phase sensitive and become degenerate at $\varphi=0$ for even $n$ and at
$\varphi=\pi$ if $n$ is odd. For low energies $\varepsilon\ll\Delta$ we estimate
$\alpha(\varepsilon)\approx\pi$ and the maximum level splitting is
$\delta\varepsilon\approx\Gamma$. We conclude that phase sensitivity survives
only in a narrow interval of order $\Delta\mu_{x0}\sim\Gamma$ around the
degeneracy points $\mu_{x0}^{\rm
  d}$, being negligible everywhere else (see Fig.~3(c)).
\begin{figure}
  \centerline{\epsfxsize=7.5cm \epsfbox{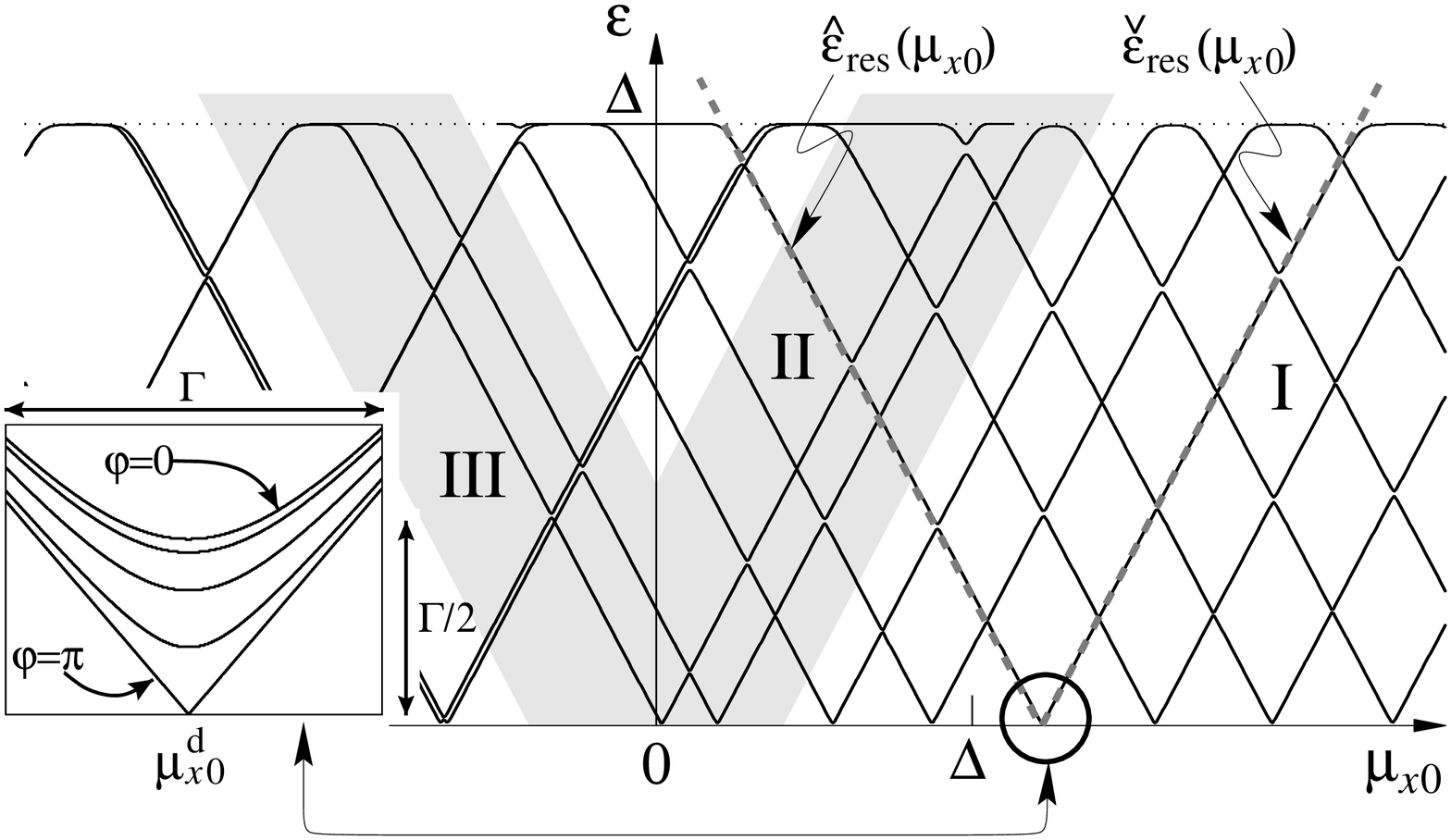}}
  \narrowtext\vspace{4mm} \caption{Discrete spectrum for a triple
    barrier generating narrow resonances (from a numerical solution of
    the quantization condition (\ref{qc})). The levels are pinned to
    the resonances; note that hole-like resonance energies
    $\check\varepsilon_{\rm res}$ grow with $\mu_{x0}$ whereas
    electronic resonance energies $\hat\varepsilon_{\rm res}$
    decrease. In region~II (III) the hole-like (electronic) levels
    pair up such that the branches of imperfect resonances carry two
    bound states. The inset shows the lowest level near the degeneracy
    point $\mu_{x0}^{\rm d}$ for different values of $\varphi$
    ($\varphi=j\pi/4$, $j=0\dots4$). }
\end{figure}
Above we have found the trapped levels belonging to a branch
$\varepsilon_{\rm res}(\mu_{x0})$ of perfect resonances at the
degeneracy points $\mu_{x0}^{\rm d}$ and far away,
$|\mu_{x0}-\mu_{x0}^{\rm d}|\gg\Gamma$. In between we can interpolate
(\ref{level1}) and (\ref{level2}) by means of the usual hyperbolic
dispersion relation and obtain (in the limit $\varepsilon\ll\Delta$;
$\alpha\approx\pi$)
\begin{eqnarray}
\label{level3} \varepsilon^{(\pm)}(\mu_{x0})=\varepsilon_{\rm res}^{\rm d}\!
\pm\!
\sqrt{(\mu_{x0}-\mu_{x0}^{\rm d})^2+[\Gamma\cos(\varphi/2)/2]^2},
\end{eqnarray}
where we have assumed that $n(\varepsilon_{\rm res}^{\rm d})$ is odd
and $\partial_{\mu_{x0}}\check\varepsilon_{\rm res}\approx
-\partial_{\mu_{x0}}\hat\varepsilon_{\rm res}\approx 1$. The result
(\ref{level3}) will be useful in our discussion of the transport
properties below.

In region~II (see Fig.\ 3(a)) the perfect hole-like resonances pair up
and collapse as $\mu_{x0}$ is lowered, thus generating imperfect
resonances, see Fig.~6. The analogous collapse of the electronic
resonances is shifted to region~III. Note that an imperfect resonance
carries {\em two} nearly degenerate levels guaranteeing that the
number of bound states remains conserved upon changing $\mu_{x0}$.

\subsection{Other Sources of Non-ideal Transmission}

The above analysis has been based on $\delta$-scatterers modelling the
effects of an insulating layer in an SIN interface. Assuming other
sources of non-ideal transmission at $\pm L/2$ we have to modify some
of the above results:
%
%
In order to describe junctions with potential steps $V_{\rm\scriptscriptstyle
S}$ at the edges we introduce the wave vector $k_{\rm nw} = \sqrt{2m(E -
V_{\rm\scriptscriptstyle S})}/\hbar$ in the normal wire and distinguish it from
the analogous quantity $k_{\rm sc} = \sqrt{2mE}/\hbar$ in the superconductor.
With $\kappa\equiv(k_{\rm sc} / k_{\rm nw} + k_{\rm nw} / k_{\rm sc})/2\geq 1$
the transmission probability takes the form $T = 2/(1+\kappa)$ (see the
Appendix). In the presence of weak resonances ($T \lesssim 1$), Eq.~(\ref{gap})
for the gaps in the Andreev spectrum reads $\delta\varepsilon\approx 4 \Lambda
\sqrt{1-T}|\cos\chi^{t_i}|/\pi$, while the results for strong resonances remain
the same (see (\ref{level1})).  A step in the effective mass at the interface
produces a similar result: with $m(|x|>L/2) = m_{\rm sc}$ and $m(|x|<L/2) =
m_{\rm nw}$ the ratio of wave vectors entering the parameter $\kappa$ takes the
form $k_{\rm sc}/k_{\rm nw}= \sqrt{m_{\rm
    sc}/m_{\rm nw}}$.  Finally, combining all three effects, a
potential step, an effective mass discontinuity, and an insulating
layer (in the form of a $\delta$-function scatterer) at the interface,
the transmission probability can be written in the form
\begin{equation}
T = \frac{4 k_{\rm sc} k_{\rm nw} m_{\rm sc} m_{\rm nw}}{(k_{\rm nw} m_{\rm sc}
+ k_{\rm sc} m_{\rm nw})^2+4 V_0^2 m_{\rm sc}^2 m_{\rm nw}^2/\hbar^4};
\label{transm}
\end{equation}
obviously, the various scattering mechanisms are non-additive and
Matthiessen's rule is not applicable.

Slightly asymmetric junctions have essentially the same properties as
the idealized symmetric ones. An NINI'N junction with insulating
layers of different transparency, $T_{\rm min}\lesssim T_{\rm max}$,
exhibits large imperfect resonances\cite{RiccoAzbel} of height
$t^2(E_{\rm res})\approx T_{\rm min}/T_{\rm max}$ instead of perfect
resonances with unit transmission.  Furthermore, $\chi^r$ is now a
continuous function of energy; the discontinuous jumps by $\pi$ are
smeared (with $\chi^r$ increasing monotonuously for $T_{-L/2} <
T_{L/2}$ and bounded for the opposite case with $T_{L/2} < T_{-L/2}$;
note that $\chi^r + \chi^{r^\prime} = \pi + 2\chi^t$,
$\chi^{r^\prime}$ the corresponding phase for a particle incident from
the left) and $\cos\beta$ changes rapidly but smoothly between $\pm 1$
at the resonance energies. Since all terms entering the quantization
condition (\ref{qc}) are only weakly affected by small asymmetries,
the above discussion of the electronic properties remains valid for
(weakly) non-symmetric junctions.

\section{Transport}

We proceed with the investigation of the transport properties of the
symmetric SINIS junctions in region I (see Fig.\ 6) and concentrate on
the situation characterized by strong resonances with $\Gamma/\Lambda
\ll 1$. Given the dependence of the quasi-particle energy
$\varepsilon$ on the phase $\varphi$, the contribution of the level to
the supercurrent follows from a simple derivative,
$I=(2e/\hbar)\partial_\varphi\varepsilon$ (a factor 2 has been
included to account for spin-degeneracy).

\subsection{Generic case}

We first consider the generic case where a level is pinned to an isolated
(electronic) resonance. In this situation we can assume $t_+\sim 1$,
$t_-\propto\Gamma/\Lambda$ and $\cos(\delta\chi^t-\alpha)
-r_+r_-\cos\beta\propto\varepsilon/\Gamma+{\rm const.}$ (see also Fig.~5).
Within this approximation $t_+t_-\propto\Gamma/\Lambda$ is small and linearly
related to $\delta\varepsilon = \max_{\varphi}[\varepsilon(\varphi)]
-\min_{\varphi} [\varepsilon (\varphi)]$ through the large slope $1/\Gamma$. We
then estimate $\delta\varepsilon\propto\partial_\varphi\varepsilon \propto
\Gamma^2$ and find that each level contributes a small supercurrent, of order
$\Gamma^2$.

\subsection{Degenerate resonances}

The phase sensitivity of the trapped levels is dramatically increased close to
degenerate resonances, see Eq.\ (\ref{level3}).  However, still assuming narrow
resonances with $\Gamma/\Lambda \ll 1$, the supercurrent in general remains
small: within our approximation the contributions arising from a pair of nearly
degenerate levels cancel each other due to the symmetry $\partial_\varphi
\varepsilon^{(+)} = -\partial_\varphi \varepsilon^{(-)}$ (a more accurate
analysis provides a residual contribution of order $\Gamma^2$, see
\cite{WendinShumeikoGJ}). Furthermore, the continuous part of the spectrum again
contributes with a term of order $\Gamma^2$. The only situation producing a
large current (of order $\Gamma$) then is realized at the special degeneracy
points produced by an (electronic) resonance crossing the Fermi level, where
$\hat{\varepsilon}_{\rm res} (\mu_{x0}^{\rm d}) \approx \check{\varepsilon}_{\rm
res} (\mu_{x0}^{\rm d}) \approx 0$, see Fig.~6. With its energy
$\varepsilon_0^{(+)}$ described by (\ref{level3}), this level carries a
nonvanishing supercurrent of magnitude
\begin{equation}
I_0= \frac{2e}{\hbar}\frac{\Gamma^2}{16}\frac{\sin\varphi}
{\sqrt{(\mu_{x0}-\mu_{x0}^{\rm d})^2+(\Gamma/2)^2\cos^2(\varphi/2)}}.
\label{current}
\end{equation}
With one channel open, we then find that a large critical supercurrent
\begin{equation}
I_{\rm c} (\mu_{x0}) =
\frac{e\Gamma}{2\hbar}\Biggl[\sqrt{1\!+\!\Big[\frac{(\mu_{x0}-\mu_{x0}^{\rm
d})}{\Gamma/2}\Big]^2}\!-\!\frac{|\mu_{x0}-\mu_{x0}^{\rm d}|}{\Gamma/2}\Biggr]
\label{critcurrent}
\end{equation}
is realized near the special values $\mu_{x0}^{\rm d}$ for the
chemical potential where ${\varepsilon}_{\rm res}^{\rm
  d}(\mu_{x0}^{\rm d}) = 0$ and for $\varphi=\pi-0$. This large
supercurrent flow quickly vanishes as $\mu_{x0}$ is tuned away from these
degeneracy points by an energy larger than the resonance width $\Gamma$, see
Fig.\ 7. Below, we refer to such a tunable SINIS junction as a {\em Resonant
Josephson-Transistor} (RJT) (alternative schemes leading to a transistor effect
make use of resonant electromagnetic pumping \cite{WendinShumeikoBratus} or
injection of quasi-particles into specific levels via multiprobe devices, see
Refs. \onlinecite{WendinShumeikoGJ,vanWees,Morpurgo,Schraepers,Baselmans}). Note
that the chemical potential $\mu_{x0}$ is not directly accessible but only
through the gate voltage $V_{\rm g}$, thus introducing the slope
$d\mu_{x0}/dV_{\rm g}$ as an additional characteristic parameter of the device.
Also, we point out that a sharp transistor effect requires the temperature to be
low, $k_{\rm\scriptscriptstyle B}T < \Gamma$.
\begin{figure}
  \centerline{\epsfxsize=7.0cm \epsfbox{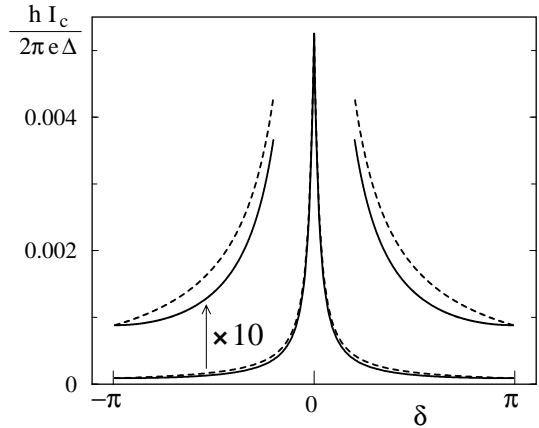}}
  \narrowtext\vspace{4mm} \caption{Critical current versus chemical
    potential $\mu_{x0} = \delta\hbar/2
    \tau|_{\varepsilon_{\rm\scriptscriptstyle F}}$ for a sharp
    resonance with $T = 0.1$ and
    $\tau|_{\varepsilon_{\rm\scriptscriptstyle F}} = 10 \hbar/\Delta$.
    The approximation (\ref{critcurrent}) (dashed line) agrees with
    the exact result obtained via the Green's function analysis at
    resonance and midway between resonances, $\delta = 0, \pm\pi$, and
    provides a good estimate in between.}
\end{figure}
The result (\ref{critcurrent}) appears to be valid in the immediate vicinity of
a resonance crossing the Fermi level. However, comparing this result with the
one derived from a full Green's function analysis \cite{Chtchelkatchev2} of the
long ($L \gg \xi$) SINIS junction with sharp resonances ($\Gamma/\Lambda \ll 1$)
one finds that the result (\ref{critcurrent}) is exact at the degeneracy points
$\mu_{x0} \approx \mu_{x0}^{\rm d}$ and, as it turns out, also midway in between
two such degeneracies; defining the parameter $\delta = 2 S |_{\varepsilon_{\rm
\scriptscriptstyle F}}/\hbar \approx 2 (\mu_{x0}-\mu_{x0}^{\rm d})
\tau|_{\varepsilon_{\rm\scriptscriptstyle F}}/\hbar +2n\pi$ the result
(\ref{critcurrent}) then is exact at all multiples of $\pi$, $\delta \approx
n\pi$ (here, $S|_{\varepsilon_{\rm\scriptscriptstyle F}}$ denotes the action for
an electron traversing the normal region $[-L/2+0, L/2-0]$ at the Fermi level
and $\tau \equiv \partial_\varepsilon S$ is the travel time). Furthermore, in
between the points $\delta = n\pi$ the deviations of (\ref{critcurrent}) from
the exact Green's function result are small, see Fig.\ 7.

\subsection{Josephson- versus Single Electron Transistor}

Next we discuss the relation between the Josephson-Transistor and the
Superconducting Single Electron Transistor (SSET)\cite{Joyez}. The
latter consists of two Josephson junctions separated by a
superconducting grain, usually referred to as the {\em island}, which
is capacitively (with capacitance $C_{\rm g}$) coupled to an external
gate electrode.  In a more figurative terminology we might call this
device a SISIS junction. While the charging energy $E_{\rm
  C}=e^2/2C_\Sigma$ tends to fix the number of Cooper pairs on the
island, the (conjugate) phase variable tends to be fixed by the
Josephson coupling energy $E_{\rm J}$ of the two junctions. For the
SSET, charging effects are dominant, $E_{\rm C} \gg E_{\rm J}$, and
thus the total capacitance $C_\Sigma$ of the island must be small (but
still\cite{Tinkham} $C_{\rm g} \ll C_\Sigma$).  Moreover we assume
$E_{\rm C} < \Delta$ and hence the ground state of the island contains
an even number of electrons at any gate voltage. Given this layout for
the device, we expect the supercurrent to be suppressed by the
`Coulomb blockade' effect\cite{Averin}. However, for specific values
$V_{\rm g}^{\rm d}$ of the gate voltage the energies of two even
charge states differing by $2e$ become degenerate. At these `Coulomb
resonances' the suppression of the Josephson current through the
island is lifted and we find a large superflow. We can control the
supercurrent by changing the gate voltage and hence the SSET
constitutes a transistor device.

The above discussion shows that the SSET and the RJT are like devices,
switching on as levels become degenerate; while for the SSET these
levels belong to fixed charge states of the island, for the RJT these
levels derive from transmission resonances. In both cases the
degeneracy is lifted by the Josephson coupling $E\cos(\varphi/2)$,
where the coupling constant $E$ is related to the transmission of the
NIS boundaries, see below. Furthermore, the SSET and the RJT exhibit
largely the same current-voltage characteristic close to these level
degeneracies,
\begin{eqnarray}
\label{sc}
I=\frac{2e}{\hbar}\frac{E^2}{4e\gamma} \frac{\sin\varphi}{\sqrt{(V_{\rm
      g}-V_{\rm g}^{\rm d})^2 +\left(E\cos(\varphi/2)/e\gamma\right)^2}},
\end{eqnarray}
where the parameters $E$ and $\gamma$ have to be specified for each
device. In fact, (\ref{sc}) yields the current through a symmetric
SSET\cite{Tinkham} if we substitute the energy parameter $E$ by the
coupling energy $E_{\rm J}$ and the dimensionless constant $\gamma$ by
the small ratio $C_{\rm g}/C_\Sigma$. Using the Ambegaokar-Baratoff
relation\cite{Ambegaokar} and the Landauer formula\cite{Landauer} we
can reexpress the coupling energy in terms of more microscopic
quantities: $E_{\rm J} = NT\Delta/4$, where $N\propto
k_{\rm\scriptscriptstyle F}^2A$ denotes the number of open channels in
the tunnel junction, $A$ denotes its area, and $T$ its transmission
coefficient. On the other hand, the RJT involves the parameters
$E=\Gamma/2$ and $\gamma=\partial\mu_{x0}/\partial eV_{\rm g}$. Again,
the resonance width is determined by geometrical quantities:
$\Gamma\approx T\Lambda/\pi$, where $T$ is the transmittivity of the
insulator layer in the triple barrier\cite{triple}.  For flat inner
barriers the quasi-classical method provides the simplification
$\Lambda\sim2(\mu_{x0}\varepsilon_{\rm\scriptscriptstyle L})^{1/2}$,
with $\varepsilon_{\rm \scriptscriptstyle L} = \hbar^2\pi^2/2mL^2$.
Note that, while charging effects dominate the physics of the SSET
these are much less relevant for the RJT as the latter involves an
`open wire' rather than a `closed island'; the adiabatic joints to the
superconducting banks provide reservoirs which effectively screen the
charge transport through the wire.

From Eq.~(\ref{sc}) we can determine the critical supercurrent $I_{\rm
  c}$ as a function of the gate voltage near $V^{\rm d}_{\rm g}$.
Apparently, $I_{\rm c}(V_{\rm g})$ defines a current peak of width
$\delta V_{\rm g} = E/e\gamma$ that attains its maximum $I_{\rm max} =
e E/\hbar$ at the degeneracy point. We estimate its slope by the ratio
$I_{\rm max}/\delta V_{\rm g}=e^2\gamma/\hbar$ carrying the dimension
of a conductance. For the SSET $\gamma\ll1$ is determined through the
small capacitance ratio $C_g/C_\Sigma$. The analogous quantity for the
RJT is of order unity, implying that the RJT shows a more prominent
slope: with $\partial \mu_{x0}/\partial eV_{\rm g} = \partial_d
\mu_{x0}/ \partial_{eV_{\rm g}} d$, and $\partial_d \mu_{x0} =
2(\varepsilon_{\rm \scriptscriptstyle F} -\mu_{x0})/d$ we have to
determine the suceptibility of the channel width $d$ with respect to
the gate voltage $V_g$. Making use of simple electrostatic
considerations one easily finds that $\partial_{\mu_{x0}} d \sim
a_{\rm\scriptscriptstyle B}/\varepsilon_{\rm \scriptscriptstyle F}$,
where $a_{\rm\scriptscriptstyle B}$ denotes the Bohr radius in the
semiconductor material \cite{GlazmanKhaetskiiE,LarkinDavies}.  Hence,
$\gamma \sim a_{\rm\scriptscriptstyle B}/d$ and with
$a_{\rm\scriptscriptstyle B}$ of order 10 nm typically we arrive at a
value of order unity for the parameter $\gamma$.

Comparable energy scales $\Delta \sim \sqrt{\mu_{x0} \varepsilon_{\rm
    \scriptscriptstyle L}}$ can be reached in realistic short SINIS
junctions ($L\lesssim \pi (\mu_{x0}/2m)^{1/2}\hbar/\Delta$).
Furthermore, we can optimize the performance of the RJT by tuning the
transparency of the insulating layers (the Dirac $\delta$-scatterers)
until the resonance width $\Gamma$ approaches the resonance spacing
$\Lambda$; such a choice of parameters produces still isolated current
peaks of maximum height. On the other hand, the maximum superflow
through the SSET scales with the number $N$ of open channels, $I_{\rm
  max} = eNT\Delta/4\hbar$, while the RJT as defined above is
generically a single channel device with a critical supercurrent
$I_{\rm max} = eT \sqrt{\mu_{x0} \varepsilon_{\rm \scriptscriptstyle
    L}}/ \hbar\pi$; going over to a many channel RJT device,
resonances from individual channels superpose at {\it different}
values for the chemical potential and hence do not add up in general
\cite{Chtchel3}. Also, a smoothing of the resonance structure has been
observed in the numerical results by Wendin {\it et al.}
\cite{WendinShumeiko}

Summarizing, superconducting transistors can be designed in terms of
`charge' (the SSET) or `phase' devices (the JT). In the charge device,
the island is separated from the superconducting leads through
insulating barriers. On the other hand, switching channels in a
perfect SNS junction one obtains a phase device with a (large)
critical current $I_c = e/(\tau_0+\hbar/\Delta)$ determined by the
effective time $\tau_0+\hbar/\Delta$ the charge needs to traverse the
normal wire\cite{Chtchel}. Such devices with perfect interfaces are
difficult to fabricate --- in reality we always have to account for a
non-ideal transmission $T<1$ through the SIN interfaces and we end up
with a Fabry-Perot type resonator device. Sequential tunneling through
the interface barriers then reduces the supercurrent by a factor $T^2$
in general. However, tuning the chemical potential to a scattering
resonance the critical current of the junction remains large, $I_c
\approx e\Gamma/2\hbar$ of order $T$ --- again, the critical current
is given by the time $\hbar/\Gamma$ the charge spends in the junction.

A second result we wish to emphasize here concerns the fact that the
critical current is carried by the lowest quasi-particle level alone.
This has been demonstrated for the SNS device in Ref.\
\onlinecite{Chtchel} and above for the SINIS junction for the case of
strong resonances (deviations from the single level result are
numerically (but not parametrically) larger away from weak
resonances). The importance to know the dispersion of this level then
provides a good {\it a posteriori} reason for studying the
quasi-particle spectrum in such junctions. Corresponding spectroscopic
experiments can be realized using multiprobe devices as proposed in
the work of van Wees {\it et al.}  \cite{vanWees}

We thank A.\ Golubov, V.\ Shumeiko, and G.\ Wendin for helpful
discussions and the Swiss National Foundation for financial support.
Work of N.M.C.\ and G.B.L.\ was partly supported by the Russian
Foundation for Basic Research under contract number RFFI-000216617.

\appendix\section*{Resonance Structures in One Dimension}

We consider a one-dimensional potential landscape $V(x)$ confined to
the interval $[-L/2, L/2]$. As a first example we concentrate on a
symmetric barrier consisting of a broad ($m\Omega^2=-\partial_x^2V$,
$\hbar\Omega\ll V(0)$) and smooth potential barrier, bounded by steps
of height $V_{\rm\scriptscriptstyle S}=V(L/2-0)-V(L/2+0)$ (see the
inset of Fig.~8). Below we make use of the wave vectors $k_{\rm sc} =
\sqrt{2mE}/\hbar$ and $k_{\rm nw} = \sqrt{2m(E-V_{\rm
    \scriptscriptstyle S})}/\hbar$ describing particles of energy $E$
in the wire close to the boundaries. The global transmission amplitude
$t\exp(i\chi^t)$ is most easily obtained by determining the
transfer-matrix\cite{RiccoAzbel} of the barrier between $-L/2-0$ and
$L/2+0$ and we find the result
\begin{eqnarray}
\label{t_step}
t\exp(i\chi^t)=\frac{t_i}{\cos(\chi^{t_i}) -i\kappa \sin(\chi^{t_i})
  \pm i r_i \sigma},
\end{eqnarray}
where $t_i\exp(i\chi^{t_i})$ and $r_i\exp(i\chi^{r_i})$ denote the
transmission and reflection amplitudes of the inner smooth barrier. In
(\ref{t_step}) we have introduced the definitions $\kappa=(k_{\rm
  sc}/k_{\rm nw}+k_{\rm nw}/k_{\rm sc})/2$ and $\sigma=(k_{\rm
  sc}/k_{\rm nw}-k_{\rm nw}/k_{\rm sc})/2$ containing the information
about the steps. In our example, $\kappa$, $\sigma$, and $t_i$ show a
weak dependence on energy and remain almost constant while
$\exp(i\chi^{t_i})$ oscillates rapidly. The derivation of
(\ref{t_step}) makes use of the relation $\chi^{r_i}-\chi^{t_i}
=\pi/2+\pi n$, $n\in\mathbf{Z}$, which follows from the unitarity of
the scattering matrix; the term $\pm i r_i\sigma$ changes sign at each
perfect resonance of $t_i$. Since the inner barrier was assumed to be
smooth, we expect $t_i$ to exhibit no resonances at all and the term
$r_i\sigma$ always carries a positive sign.

Next, we show that (\ref{t_step}) qualitatively reproduces the
transmission amplitude shown in Fig.~2. We definine the function
$z=1/t\exp(i\chi^t)$, whose four arguments $\chi^{t_i}$, $\kappa$,
$\sigma$, and $t_i$ depend on the energy. Since the inner scattering
phase $\chi^{t_i}$ is the most energy sensitive argument we minimize
$|z|$ with respect to $\chi^{t_i}$ (keeping $\kappa$, $\sigma$, and
$t_i$ fixed at a given energy $E$). Thus we obtain the stationary
phases $\chi^{t_i}_{\rm min}(E)$ belonging to the minima of $|z|$.
Moreover, we can introduce the complex valued function $z_{\rm
  min}[E]=z[\chi^{t_i}_{\rm min}(E),\kappa(E),\sigma(E),t_i(E)]$,
which is roughly constant when compared to $z[E]$. We can estimate the
resonance energies by solving the equation $z[E_{\rm res}]=z_{\rm
  min}[E_{\rm res}]$ for $E_{\rm res}$. Apparently, $z[E]$ describes
an ellipse in the complex plane with half axes $1/t_i$ and
$\kappa/t_i$. The center of this ellipse is shifted away from the
origin by $ir_i\sigma/t_i$.
\begin{figure}[htb]
  \centerline{\epsfxsize=8.5cm \epsfbox{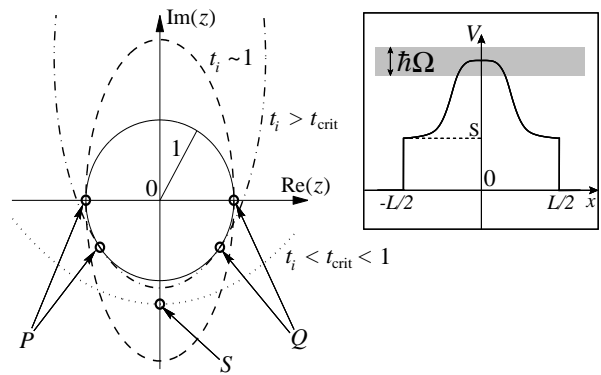}}
  \narrowtext\vspace{4mm} \caption{Complex transmission amplitude
    through a smooth barrier bounded by steps. With decreasing energy
    $E$ the inverse $z=1/t\exp(i\chi^t)$ moves counterclockwise on an
    ellipse, the center of which is positioned at $ir_i\sigma/t_i$ and
    shifts up the imaginary axis.  As $t_i$ drops below $t_{\rm crit}$
    the perfect resonances at $P$ and $Q$ collapse at the bottom of
    the ellipse ($P\rightarrow S\leftarrow Q$) and transform into an
    imperfect (double) resonance.}
\end{figure}
Fig.~8 illustrates the behavior of $z_{\rm
  min}$ when $r_i$ grows from 0 to 1 in the intermediate region
$E\in[V(0)-\hbar\Omega/2,V(0) +\hbar\Omega/2]$ (with a smooth inner
barrier, $r_i$ is strictly monotonous, $r_i\approx0$ for
$E>V(0)+\hbar\Omega/2$ and $r_i\approx1$ for $E<V(0)-\hbar\Omega/2$).
The distance from the origin to the bottom of the ellipse (point $S$)
is always extremal. At high energies where $1\approx
t_i>1/\kappa=t_{\rm crit}$, we observe {\em perfect} resonances: with
$|z_{\rm min}|=1$ these resonances are realized at the symmetric
points $P$ and $Q$ where the ellipse touches the unit circle, see
Fig.~8. As $t_i$ decreases the perfect resonances approach each other
($P$ and $Q$ move towards $S$) and merge with $S$ as
$t_i{\scriptscriptstyle ^{\searrow}} t_{\rm crit}$. At energies
$E\lesssim V(0)$ ($\Rightarrow t_i<t_{\rm crit}$) the reflection
coefficient $r_i$ drops to zero, and $S$ becomes the closest point to
the origin with $|z_{\rm min}|=(\kappa- \sigma r_i) /t_i>1$; the
resonances then have paired up and have become {\em imperfect}, with a
height decaying rapidly with decreasing energy $E$.

Physically, this resonance structure originates from the interplay of
two competing transparencies. The transmission probability through the
potential steps is almost constant and much smaller than the
transparency of the inner barrier for $E\gg V(0)$. In this regime the
particles propagate freely and are reflected by the steps alone; thus
we observe perfect resonances, and the shape of the smooth inner
barrier only influences the resonance spacing. On the other hand, for
$E\ll V(0)$ the inner barrier behaves like a hard wall while the steps
are comparatively transparent. The global transmission through the
barrier bounded by the steps is suppressed by the tunneling through
the classically inaccessible region. The resonances of the subsystems
consisting of a step and a potential hill account for the global
imperfect resonances. Note that the barrier bounded by steps generates
pronounced resonances only if $V_{\rm \scriptscriptstyle S}\gtrsim
0.7~V(0)$.
\begin{figure}
  \centerline{\epsfxsize=8.5cm \epsfbox{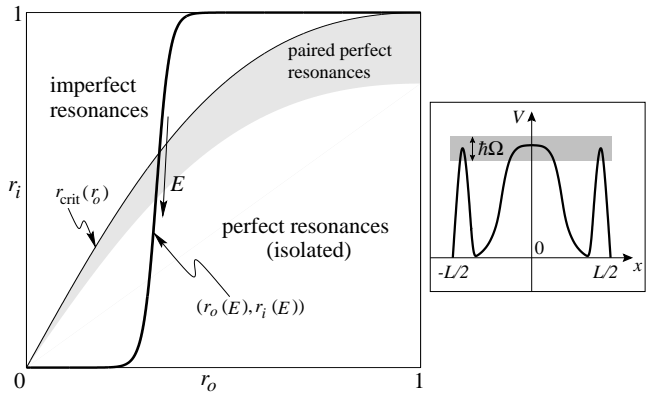}}
  \narrowtext\vspace{4mm} \caption{The solid curve in the $(r_o,r_i)$
    plane characterizes the energy dependence of the inner and the
    global reflection amplitudes in a symmetric triple barrier as
    shown in the inset. The domains of perfect and imperfect
    resonances are separated by the curve $r_{\rm crit}(r_o)$.  In an
    narrow interval $\sim\hbar\Omega$ around $V(0)$ the reflection
    coefficent $r_i^2$ drops from $1$ to $0$; in this energy range
    $r_o$ can be considered as a constant.}
\end{figure}
In a second step we investigate the transmission amplitude through
three arbitrary symmetric potential barriers, i.e., a triple barrier,
see Fig.~9. The inner barrier is again smooth, whereas the two outer
barriers (characterized by the transmission and reflection amplitudes
$t_o\exp(i\chi^{t_o})$ and $r_o\exp(i\chi^{r_o})$) are assumed to be
equal. Thus the global transmission amplitude takes the form
\begin{eqnarray}
\label{t_triple} te^{i\chi^t}=\frac{t_i
t_o^2e^{i(\chi^{t_i}+2\chi^{t_o})}}{1+2r_ir_o
e^{i(\chi^{t_i}+\chi^{t_o})}- r_o^2 e^{i2(\chi^{t_i} +\chi^{t_o})}}.
\end{eqnarray}
In order to determine the resonance structure we evaluate the extrema
of the modulus of the denominator in (\ref{t_triple}) with respect to
$\chi^{t_i}+\chi^{t_o}$ (we fix $t_i$ and $t_o$, which are less
sensitive to changes in energy). A direct calculation yields the two
extremal conditions $\cos(\chi^{t_i}+\chi^{t_o}) = -r_i(1+r_o^2)/2r_o$
(I) and $\sin(\chi^{t_i}+\chi^{t_o}) =0$ (II). Elementary
manipulations show that the transmission amplitude exhibits {\em
  perfect} resonances in the first case. But condition~I can only be
satisfied as long as $r_i \leq r_{\rm crit}(r_o)=2r_o/(1+r_o^2)$,
i.e., for a reasonably transparent inner barrier (see Fig.~9). The
extremal condition~II requires a second distinction. Condition~II(a)
reads $\chi^{t_i}+\chi^{t_o}=2\pi n$, $n\in\mathbf{Z}$. The critical
points of this type always belong to local minima of $t$. Finally, we
discuss condition~II(b), $\chi^{t_i}+\chi^{t_o}=\pi+2\pi n$,
$n\in\mathbf{Z}$, which refers to local minima when $r_i<r_{\rm crit}$
and characterizes {\em imperfect} resonances in the regime $r_i>r_{\rm
  crit}$. Since the internal barrier was assumed to be smooth, the
regime of perfect resonances lies at high energies $E>V(0)+
\hbar\Omega/2$, where $r_i\approx 0$ and $r_o\gg r_i$. At low energies
$E<V(0)- \hbar\Omega/2$, the resonances are always imperfect. Within
the crossover region the perfect resonances attract each other
pairwise and collapse to become imperfect as the energy decreases (see
Fig.~2).

Other sources of resonances are discontinuities of the force
$\partial_x V$ and effective mass steps. Technically they can be
treated like Dirac $\delta$-scatterers and potential steps,
respectively, as can be easily checked making use of the
transfer-matrix formalism.


\end{multicols}

\begin{thebibliography}{10}
\bibitem{conv} Note that in Ref.~\onlinecite{Chtchel} basis states
have been chosen where the phases vanish at the center of the
junction.
\bibitem{sepNS} It is convenient to separate the effects of normal and
Andreev scattering through a thin normal metallic layer of width
$\delta \rightarrow 0$.
\bibitem{gamma} Note that in our convention $\Gamma$ is twice as large
as in Ref.~\onlinecite{WendinShumeiko}.
\end{thebibliography}
\end{document}